\documentclass[twocolumn,preprintnumbers,prl,nobibnotes,nofootinbib]{revtex4}

\usepackage{amsmath}
\usepackage{graphicx}

\newcommand{\gev}{{\ \rm GeV}}
\newcommand{\tev}{{\ \rm TeV}}

\def\mpl{\ifmmode \overline M_{Pl}\else $\bar M_{Pl}$\fi}
\newcommand{\ie}{{\it i.e.}}

\newskip\zatskip \zatskip=0pt plus0pt minus0pt
\def\matth{\mathsurround=0pt}

\def\gsim{\mathrel{\mathpalette\atversim>}}
\def\atversim#1#2{\lower0.7ex\vbox{\baselineskip\zatskip\lineskip\zatskip
  \lineskiplimit 0pt\ialign{$\matth#1\hfil##\hfil$\crcr#2\crcr\sim\crcr}}}

\begin{document}

\author{JoAnne L. Hewett} \email{hewett@slac.stanford.edu}
\author{Ben Lillie} \email{lillieb@slac.stanford.edu}
\author{Thomas G. Rizzo} \email{rizzo@slac.stanford.edu}
\affiliation{Stanford Linear Accelerator Center, 2575 Sand Hill
  Rd. Menlo Park, CA 94025}
\thanks{Work supported by the
  Department of Energy
  Contract DE-AC02-76SF00515.}

\title{Black holes in many dimensions at the LHC: testing critical string theory}

\date{\today}
\preprint{SLAC-PUB-11024}

\begin{abstract}
We consider black hole production at the LHC in a generic scenario with
many extra dimensions where the Standard Model fields are confined to a
brane. With $\sim 20$ dimensions the hierarchy problem is shown to be naturally
solved without the need for large compactification radii. We find that in such
a scenario the properties of black holes can be used to determine the number
of extra dimensions, $n$. In particular, we demonstrate that measurements of
the decay distributions of such black holes at the LHC can determine if $n$ is
significantly larger than 6 or 7 with high confidence, and thus can probe one
of the critical properties of string theory compactifications.
\end{abstract}

\maketitle



One of the most difficult questions facing theoretical high-energy
physics is how to consistently combine General
Relativity with Quantum Mechanics, as naive quantization produces
unrenormalizable divergences. This issue is exacerbated by the
hierarchy problem, which asks why the electroweak scale,
$M_{wk}\sim$ TeV, is so small compared with the (reduced) Planck
scale, $\mpl \sim$ a few $10^{18}$ GeV, which is associated with the
energy at which non-renormalizable Einstein gravity becomes strong.
It appears that resolution of these puzzles may require a complete
theory of quantum gravity.

In recent years it has been proposed that the fundamental scale of
gravity might not be $\mpl$, but rather $M_* \sim \tev$
\cite{Antoniadis:1990ew,Arkani-Hamed:1998rs}. There is then no large
hierarchy between the gravitational and electroweak scales. In this
scenario, the observed weakness of gravity results from the presence
of extra dimensions with large radii. In the simplest picture,
gravity is able to propagate in all $D$ dimensions, but the Standard 
Model (SM) fields are restricted to a $3+1$ dimensional ``brane''. The 
strength of gravity at long distances is then diluted by the volume of the
extra dimensions.  Here, we examine a scenario where the 
number of extra dimensions is large.  In this case, as we will see below, 
additional hierarchies do not arise between $M_*$ and the size of the
additional dimensions.  In particular, we examine the
properties of black hole (BH) production and decay at the LHC with
different numbers of extra dimensions and show that the number of
additional dimensions $n$ can be determined at high confidence, in
particular when $n$ is large.  Our results hold in the {\it generic case} 
where the size of the BH is much less than the curvature of the additional 
dimensions, and where the SM is confined to a 3-brane.

As of now, the best candidate for a complete theory of quantum
gravity is (critical)
string theory (CST), which reduces to Einstein gravity at low
energies and allows for the computation of finite $S-$matrix
amplitudes. For CST to be a consistent theory there are three
essential ingredients: ($i$) the fundamental objects of the theory
are no longer point-like and must have a finite size of order
$M_{s}$, the string scale; ($ii$) supersymmetry must be a good
symmetry, at least at scales $\gtrsim M_{s}$; ($iii$) space-time
must be ten or eleven dimensional, (\ie, $D=4+n=10$, if the string
coupling is perturbative, $D=11$ if it is non-perturbative), with
the additional dimensions being compactified at a radius $R_c\gtrsim
1/M_{s}$. Most research in string theory so far has focused on
critical string theories, where the world-sheet anomalies are
automatically canceled. It is precisely this anomaly cancelation
that requires $D=10$. However, there are consistent non-critical
backgrounds of string theory in arbitrary numbers of dimensions.
Here, the anomalies are canceled by solving the equations of motion
taking into account the tree level moduli potential as well as
contributions to the equations of motion from other sources such as
fluxes, orientifolds, and branes \cite{Myers:1987fv}. In either
case, the common expectation is that $M_{s}$ is slightly below or
equal to $\mpl$ which would imply that the predictions of CST are
difficult to test directly. Currently there is no evidence for any
of these basic assumptions. If indeed $M_{s} \sim \mpl$ it may be
that CST can never be directly tested in laboratory experiments.
Furthermore, even if supersymmetry and/or extra dimensions were
discovered in future experiments, this would be no guarantee that
CST represents the correct theory of nature.

We will show in this paper that the number of compactified 
large dimensions can be determined from black hole production at the LHC.
This would provide a probe of classes of CST models.  Specifically,
if $n > 6(7)$ is measured with high confidence then present CST 
compactifications would be tested.
As a proof of principle for our
proposal, we will show that there exists a region in the 
parameter space where we can experimentally exclude
the case $n\le 6(7)$ at $5\sigma$ significance.

For purposes of demonstration, we perform our calculations in the
the large extra dimensions picture of Arkani-Hamed, Dimopoulos and
Dvali (ADD) \cite{Arkani-Hamed:1998rs}.  We emphasize that our results
are {\it general} and we only use ADD as a calculational framework.
Here, $M_*$ and $\mpl$ are related
by $\mpl^2=V_nM_*^{n+2}$, where $V_n$ is the volume of the $n$
compactified large dimensions. 
For simplicity in what follows, we will assume that this
$n-$dimensional space is compactified on a torus of equal radii so
that $V_n=(2\pi R_c)^n$, where $R_c$ is the compactification radius.
Given $\mpl$ and $M_*\sim$ a few$\tev$, $R_c$ becomes completely
fixed by the relation above. Note that the case $n=1$ is excluded
while $n=2$ with low $M_*$ is disfavored by current data
\cite{Hewett:2002hv}. For the case of a torus, the graviton has
Kaluza-Klein(KK) excitations $h_{\mu\nu}^{(n)}$, with masses given
by $M_n^2=${\bf n}$^2/R_c^2$, where {\bf n} labels a set of
occupation numbers. The KK graviton couplings to the Standard Model
(SM) fields are described by the stress-energy tensor $T^{\mu\nu}$,
given in $D$ dimensions by ${\cal L}=-\sum_n
h_{\mu\nu}^{(n)}T^{\mu\nu}/M_*^{1+n/2}$. This scenario has three
distinct experimental signatures which have been studied in some
detail in the literature: ($i$) missing energy events associated
with KK graviton emission in the collisions of SM fields; ($ii$) new
contact interactions associated with spin-2 KK exchanges between SM
fields \cite{Han:1998sg}; ($iii$) black hole production in particle
collisions \cite{Dimopoulos:2001hw,Giddings:2001bu}.

\begin{figure}
  \includegraphics[height=6cm,width=4cm,angle=-90]{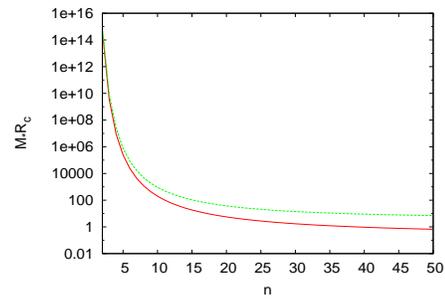}\\
  \caption{$M_* R_c$ as a function of $n$ for $M_* = 1$ TeV for a torus (solid) and sphere (dashed) compactifications.}\label{fig:rcvsn}
\end{figure}

Is there any guide as to what values of $n>6(7)$ we should consider?
For $n\leq 6$ it is well known that the hierarchy problem
is {\it not} truly solved. Although we have reduced $M_*$ to a few
TeV, $M_*R_c \gg 1$, as seen in Fig \ref{fig:rcvsn}. By contrast,
with $n$ large we could have $M_*R_c \lesssim 10$. Note that, if
$M_*R_c <1$ the theory would lose its predictive power since the
compactification scale is above the cutoff. To obtain the
interesting range of compactification radii, $1\lesssim M_*
R_c\lesssim 10$, requires $17 \lesssim n \lesssim 39$, hence we will
focus on this set of values in what follows. If the compactification
topology is a sphere, rather than a torus, this changes to $n\gsim
30$, as seen in Fig \ref{fig:rcvsn}. It is important to notice that
this model does solve the hierarchy problem for large $n$, but this
would lie outside the realm of CST. Note that some other
modifications of the compactification geometry can obtain $R_cM_*
\lesssim 10$ \cite{Kaloper:2000jb}. For such large values of $n$ the
Kaluza-Klein masses are at the$\tev$ scale.  For example, in the ADD
case, since each graviton KK state is
coupled with 4 dimensional Planck strength, $\mpl$, to the SM
fields, it is clear that this significantly weakens the KK
contributions to the processes ($i$) and ($ii$) above.  Thus, no
meaningful collider constraints would be obtainable; this may also
happen in the generic model we consider here.  For example, in ADD with 
$n=2$, precision measurements at the International Linear Collider at
$\sqrt{s} = 1 \tev$ will be sensitive to $M_* \lesssim 10 \tev$,
while with $n=21$, this drops to $M_* \lesssim 1 \tev $. Thus for
reasonable values of $M_*$ the {\it only} signal for large $n$ in
ADD is black hole production.

\begin{figure}
  \includegraphics[height=8cm,width=4.3cm,angle=-90]{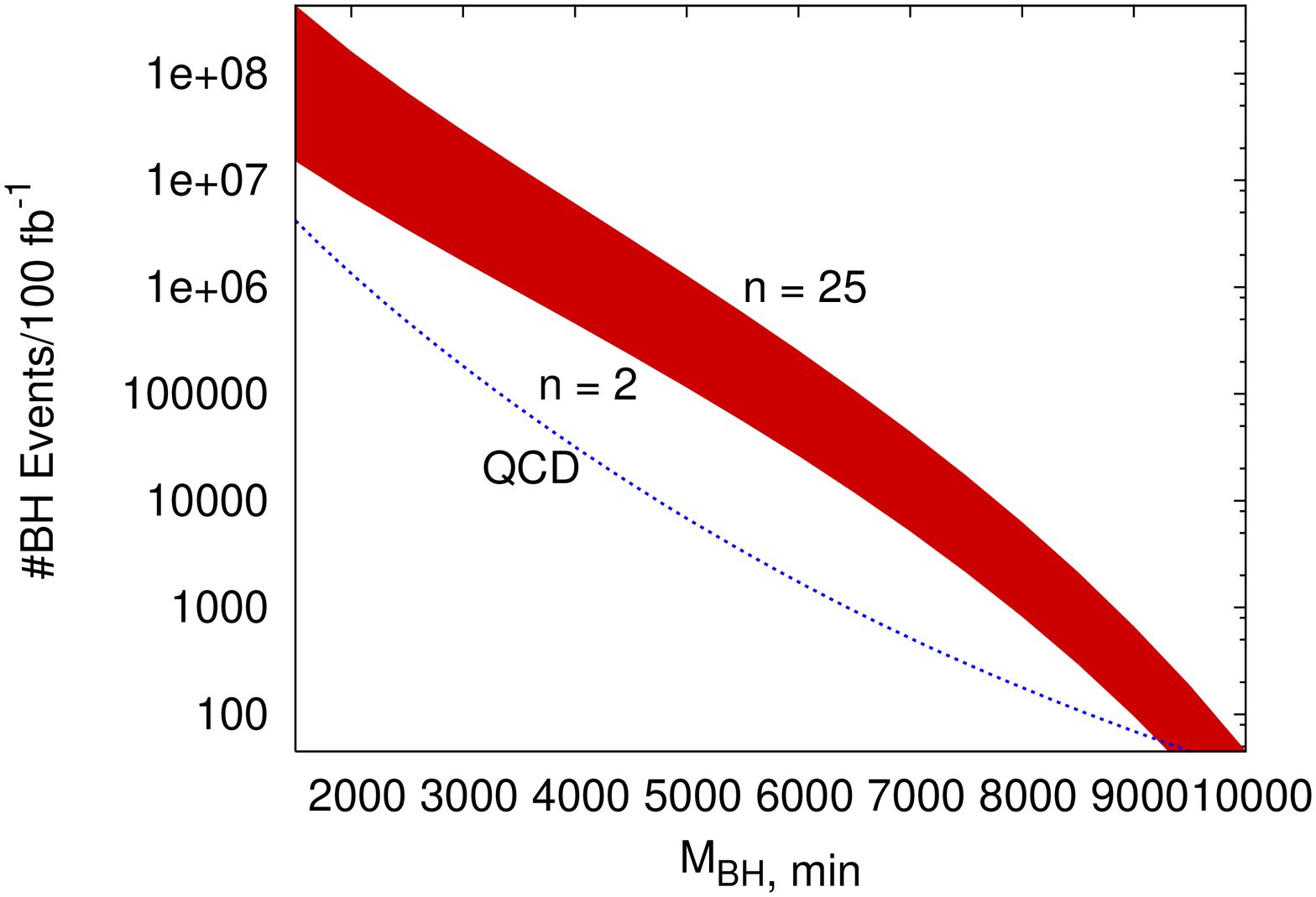}
  \includegraphics[height=8cm,width=4.3cm,angle=-90]{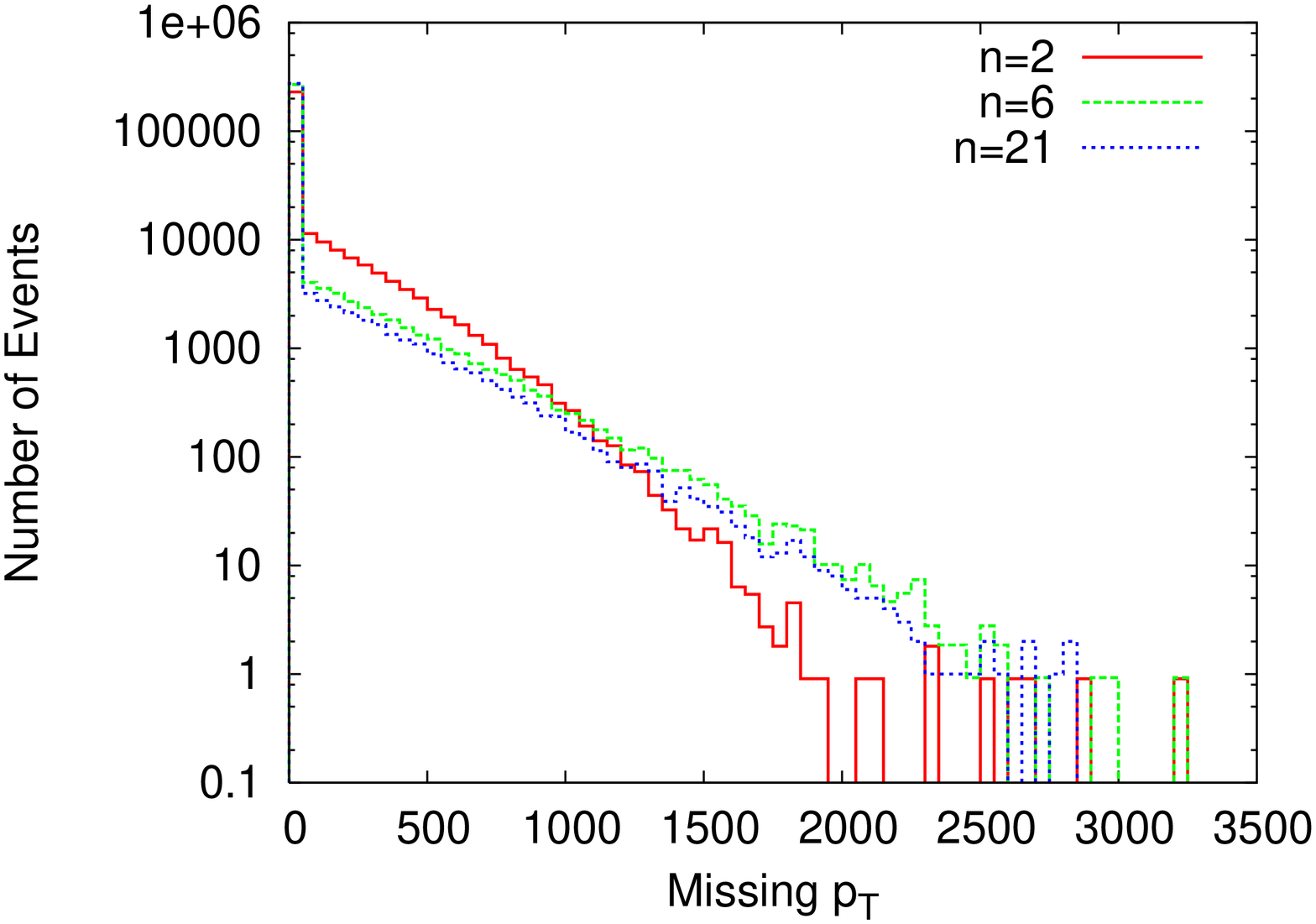}
  \caption{Top panel: Cross-section for production of black holes with mass
  $M>M_{BH, min}$ with $M_*= 1.5 \tev$, for $n=2$(bottom) to $25$(top) of the band. Also shown is the
  QCD dijet cross-section for dijet invariant mass $M\ge M_{BH, min}$, and $|\eta| < 1$.
  Bottom panel: $\not\! p_T$ distribution of BH events passing cuts described in the text for $M_* = 1 \tev$
  and $n=2,6,21$.}\label{fig:xsec}
\end{figure}

We now investigate BH production at the LHC in detail; for previous
studies see \cite{Harris:2004xt}. When $\sqrt s \gtrsim M_*$ BHs are
produced with a geometric (subprocess) cross section, $\hat \sigma
\simeq \pi R_s^2$.  We expect this to hold in all models which satisfy
our assumptions. Here $R_s$ is the Schwarzschild radius
corresponding to a BH of mass $M_{BH}\simeq \sqrt {\hat s}$. $R_s$
is given by \cite{Kanti:2004nr}
\begin{equation}
M_* R_s=\Bigg[\frac{\Gamma({\frac{n+3}{2}})}{(n+2)\pi^{(n+3)/2}}
\frac{M_{BH}}{M_*}\Bigg]^{1/(n+1)}\,.\label{eq:rs}
\end{equation}
Note that $\hat \sigma \sim n$ for large n. Numerical simulations
and detailed arguments have shown that the geometric cross section
estimate is good to within factors of a few \cite{Giddings:2004xy}.
The total number of BH events at the LHC with invariant mass above
an arbitrary value $M_{\rm BH, min}$ is shown in Fig.
\ref{fig:xsec}. The scale of the total inclusive BH cross-section,
$\sim 100$ pb, is huge compared to that which is typical of new
physics processes, $\lesssim 1$ pb. Thus, over much of the parameter
space the LHC will be producing over a million BH events per year.
This high rate means that there will be tremendous statistical
power, and essentially all measurements will be systematics limited.

The semiclassical treatment, used here and in all previous studies
\cite{Kanti:2004nr}, may recieve potentially large corrections from
two sources: ({\it i}) distortions from the finite compactification
scale as $R_s$ approaches $R_c$, and ({\it ii}) quantum gravity.
Case ({\it i}) is easily controllable. We know that in 5 dimensions
the critical point for instabilities due to finite compactification
is $(R_s/R_c)^2 \approx 0.1$ \cite{Gregory:1993vy}. For LHC energies
we always have $(R_s/R_c)^2 \ll 0.1$, so these corrections are
negligible. In more dimensions the ratio of the volume of a BH with
fixed $R_s$ to the volume of the torus with fixed $R_c$ drops
rapidly with $n$, so we expect the corrections to be even smaller.
Case ({\it ii}) is more problematic; we estimate the quantum gravity
effects by looking at the corrections from higher curvature terms in
the action, {\it e.g.}
\begin{gather}
S = \frac{M_*^{D-2}}{2}\int d^D x\ \left(R +
\frac{\alpha_1}{M_*^2}{\mathcal L}_2 +
\frac{\alpha_2}{M_*^4}{\mathcal L}_3 +
\ldots\right).\label{eq:action}
\end{gather}
Here $R$ is the Ricci scalar, and ${\mathcal L}_i$ is the $i$th
order Lovelock invariant, with ${\mathcal L}_2$ being the
Gauss-Bonnet term \cite{lovelock:498}. This equation also defines
our convention for the fundamental scale $M_*$.\footnote{We note
that this is related to the other definitions in the literature by
$M_* = (8\pi)^{-\frac{1}{n+2}}M_{\rm DL}$\cite{Dimopoulos:2001hw} $=
[2(2\pi)^n]^{-1\frac{1}{n+2}} M_{\rm GT}$\cite{Giddings:2001bu} $=
(2\pi)^{-n/(2+n)}M_D$, as used in Giudice {\it et. al}
\cite{Han:1998sg}.} Schwartzchild solutions are known for arbitrary
values of the $\alpha_i$ \cite{Whitt:1988ax}. If we assume that the
higher curvature terms are radiatively generated, and hence each
$\alpha_i$ is the $i$th power of an expansion parameter $\alpha$ (as
occurs in string models \cite{Zwiebach:1985uq}), we find that
$\alpha D^2 \le 1$. For $\alpha$ of this size we find that the
corrections are always less severe as $n$ increases, with a $\sim
20\%$ correction to $R_s$ for $n=20$. This does not qualitatively
affect our conclusions here; for a more detailed study of these
corrections, see \cite{hlr-progress,Rizzo:2005fz}.

We now come to the crucial question, is there any property of the
produced black holes that can resolve the number of dimensions? The
cross-section is $n$-dependent, but the overall scale is set by
$1/M_*^2$, so one would first have to measure $M_*$ independently to
good accuracy to obtain any resolution on $n$. Cross section ratios
at different BH masses could be used, however, the range of energies
that are clearly in the geometric regime and accessible to the LHC
is not likely to be large. This leads us to the decay properties of
black holes. One generically expects that black holes produced at
colliders are formed in highly asymmetric states, with high angular
momentum, and possibly a non-zero charge. However, they quickly shed
their charge and angular momentum by emitting bulk graviton modes
and soft brane modes, and relax to a simple Schwartzchild state;
their decay then proceeds primarily by thermal emission of Hawking
radiation \cite{Kanti:2004nr} until $M_{BH} \sim M_*$, where quantum
gravity effects will mediate the final decay. The Hawking
temperature is given by
\begin{gather}
T_H = \frac{(n+1)M_*}{4\pi}
\Bigg[\frac{\Gamma({\frac{n+3}{2}})}{(n+2)\pi^{(n+3)/2}}
\frac{M_{BH}}{M_*}\Bigg]^{-1/(n+1)}.
\end{gather}
From this we can see that, at fixed $M_{BH}$, higher dimensional BHs
are hotter. Since the average multiplicity goes inversely with the
temperature, a low dimensional BH will emit many quanta before
losing all of it's energy. By contrast, the decay of a high
dimension BH will have fewer final state particles, and each emitted
quanta will carry a larger fraction of the BH energy. We will use
this difference to obtain experimental resolution on $n$. It was
seen in \cite{Harris:2004xt} that for $n\le 6$ an error of $\pm
0.75$ could be obtained. However, as $n$ gets large the BH
properties at adjacent $n$ converge, so it is {\it a-priori} unclear
at what level $n$ can be determined, if at all, in this case.

The previous argument suggests we examine the final state
multiplicity, or the individual particle $p_T$ distributions as a
probe of $n$. The multiplicity is affected by two major sources of
uncertainty: (a) contributions from initial and final state
radiation that produce additional jets, and (b) the details of the
final quantum gravity decay of the BH are unknown. In what follows
we will assume that this {\it remnant} decay is primarily 2-body.
However, this is clearly model-dependent; we prefer observables that
are independent of this assumption, disfavoring the multiplicity. By
contrast, the $p_T$ spectra of individual particles, particularly at
high-$p_T$, will be mostly sensitive to the {\it initial}
temperature of the BHs. There are many such distributions that one
could consider. In particular, one would like to examine all
possible distributions and see that the candidate BH states are
coupling equally to each SM degree of freedom, verifying that these
are gravitational phenomena \cite{hlr-progress}. For illustration we
will focus here on the $\not\! p_T$ and individual jet $p_T$
distributions for the BH final state.

To calculate these distributions, we have simulated BH events using
a modified version of CHARYBDIS \cite{Harris:2003db}, linked to
PYTHIA \cite{Sjostrand:2000wi}. First, a large sample of BHs with
masses above a critical value $M_{\rm min} = M_*$ is generated. From
these we select events by cutting on the reconstructed invariant
mass, $M_{\rm inv}$ of the event, defined by summing over all
visible final state particles or jets with rapidity $|\eta|<3$, and
with $p_T \ge 50 \gev$. We would like to select events where the BH
mass is large enough that the event is in the geometrical regime,
and quantum gravity corrections are small. To do this, one would
need to extract from the data an estimate of the size of $M_*$.
While we have no fundamental model for the quantum gravity effects
near threshold, we can assume that there will be a turn-on for BH
production near $M_*$, and the cross-section will then asymptote to
the geometric value. While this will not lead to a precision
determination of $M_*$, it can clearly be used to set an optimum cut
on $M_{\rm inv}$. In the context of a particular model of the
threshold based on the action (\ref{eq:action}), we find that
$M_{\rm inv} \ge 2M_*$ is a reasonable cut \cite{hlr-progress}. We
include initial-state radiation in the simulations, since that can
lead to a contamination of lower $\sqrt{\hat s}$ events in our
sample. In the case of jets, for simplicity we turn off
hadronization, and simply look at the parton-level characteristics.

\begin{figure}
  \includegraphics[width=6cm,angle=-90]{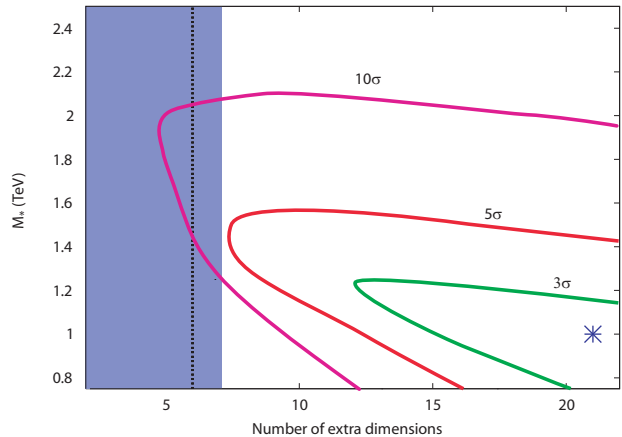}\\
  \caption{Exclusion curves in the $(n,M_*)$ plane, assuming the data
  lies at the point $(21,1\tev)$. Points outside the curves are excluded
  at 3, 5, or 10$\sigma$.}\label{fig:sigmapt}
\end{figure}

To be specific, we generate a ``data'' set of $\sim 300k$ events
with $n=21$ and $M_* = 1 \tev$. We use this size sample as a
conservative lower estimate of BH production. If the cross section
is within an order of magnitude of that in Fig. \ref{fig:xsec}, the
LHC will collect many millions of events, giving an increase in
statistical power over that presented here.  Alternatively, if we 
employed a stiffer cut on the lower value of $M_{inv}$, this would 
yield a lower statistical sample, similar to the size of $300k$ events 
considered here, and we would expect our results to then
qualitatively hold in this case as well.
These ``data'' events
are then compared to a number of template sets of events. We then
ask at what confidence the template can be excluded by performing a
$\chi^2$ test using only the resulting $\not\! p_T$ distribution
(shown in Fig. \ref{fig:xsec}). We examine the range $2\le n\le 21$,
and $0.75 \le M_* \le 5 \tev$. The lower bound on $M_*$ comes from
non-observation at the Tevatron and cosmic rays
\cite{Anchordoqui:2003jr}, while the upper bound is set by demanding
that the LHC be able to collect at least $50k$ events given the
cross-section uncertainties. We then determine whether the CST
region can be probed at high confidence within this scenario. 
For this test case, we
find at least a $5\sigma$ exclusion for the entire CST region using
the $\not\! p_T$ distribution alone, or $\sim 40\sigma$ using the
jet-$p_T$ spectrum. Though the statistical power in jets is much
higher, it suffers from more systematic uncertainties. Fig.
\ref{fig:sigmapt} shows the 3, 5, and 10$\sigma$ exclusion contours
in the $(n,M_*)$ plane obtained using the $\not\! p_T$ distribution
for this test case. If the LHC collects a few million events rather
than the $300k$ sample used here, simple scaling tells us that the
$5\sigma$ curve excludes $n\le 20$, and the $10\sigma$ curve
excludes $n\le 11$.

We have shown that the CST region can be excluded within this scenario 
if $n=21$. What
about other values of $n$? On changing the number of dimensions used
in generating the``data'', we find that for any $n\ge 15$ the CST
region can be excluded by at least $5\sigma$, with $300k$ events. We
would, of course, like to know in what region of the parameter space
this type of definitive test can be performed. A more detailed study
of the parameter space is in progress \cite{hlr-progress}.

In conclusion, we have shown that if there exist many$\tev$ sized
extra dimensions with the SM fields confined to a 3-brane, 
then there
exists an observable that can probe classes of critical string theory
models.

\begin{acknowledgments}
The authors would like to thank N. Arkani-Hamed, C. Berger, D. J.
Copeland, S. Dimopolous, G. Landsberg, L. McAllister, P. Richardson,
E. Silverstein, B. Webber, R. Wijewardhana, and T. Wiseman for many
helpful discussions.
\end{acknowledgments}


\begin{thebibliography}{99}



\bibitem{Antoniadis:1990ew}
I.~Antoniadis,
Phys.\ Lett.\ B {\bf 246}, 377 (1990).
  J.~D.~Lykken,
  Phys.\ Rev.\ D {\bf 54}, 3693 (1996)
  [arXiv:hep-th/9603133].
  E.~Witten,
  Nucl.\ Phys.\ B {\bf 471}, 135 (1996)
  [arXiv:hep-th/9602070].


\bibitem{Arkani-Hamed:1998rs}
N.~Arkani-Hamed, S.~Dimopoulos and G.~R.~Dvali,
Phys.\ Lett.\ B {\bf 429}, 263 (1998) [arXiv:hep-ph/9803315].

\bibitem{Myers:1987fv}
  R.~C.~Myers,
  Phys.\ Lett.\ B {\bf 199}, 371 (1987).
  A.~Maloney, E.~Silverstein and A.~Strominger,
  arXiv:hep-th/0205316.

\bibitem{Cullen:2000ef}
  E.~Dudas and J.~Mourad,
  Nucl.\ Phys.\ B {\bf 575}, 3 (2000)
  [arXiv:hep-th/9911019].
S.~Cullen, M.~Perelstein and M.~E.~Peskin,
Phys.\ Rev.\ D {\bf 62}, 055012 (2000) [arXiv:hep-ph/0001166].

\bibitem{Randall:1999ee}
  L.~Randall and R.~Sundrum,
  Phys.\ Rev.\ Lett.\  {\bf 83}, 3370 (1999)
  [arXiv:hep-ph/9905221].

\bibitem{Davoudiasl:2003me}
  H.~Davoudiasl, J.~L.~Hewett, B.~Lillie and T.~G.~Rizzo,
  Phys.\ Rev.\ D {\bf 70}, 015006 (2004)
  [arXiv:hep-ph/0312193].

\bibitem{Hewett:2002hv}
J.~Hewett and M.~Spiropulu,
Ann.\ Rev.\ Nucl.\ Part.\ Sci.\  {\bf 52}, 397 (2002)
[arXiv:hep-ph/0205106].

\bibitem{Han:1998sg}
T.~Han, J.~D.~Lykken and R.~J.~Zhang,
Phys.\ Rev.\ D {\bf 59}, 105006 (1999) [arXiv:hep-ph/9811350].
J.~L.~Hewett,
Phys.\ Rev.\ Lett.\  {\bf 82}, 4765 (1999) [arXiv:hep-ph/9811356].
G.~F.~Giudice, R.~Rattazzi and J.~D.~Wells,
Nucl.\ Phys.\ B {\bf 544}, 3 (1999) [arXiv:hep-ph/9811291].
E.~A.~Mirabelli, M.~Perelstein and M.~E.~Peskin,
Phys.\ Rev.\ Lett.\  {\bf 82}, 2236 (1999) [arXiv:hep-ph/9811337].
  T.~G.~Rizzo,
  Phys.\ Rev.\ D {\bf 59}, 115010 (1999)
  [arXiv:hep-ph/9901209].

\bibitem{Dimopoulos:2001hw}
  P.~C.~Argyres, S.~Dimopoulos and J.~March-Russell,
  Phys.\ Lett.\ B {\bf 441}, 96 (1998)
  [arXiv:hep-th/9808138].
S.~Dimopoulos and G.~Landsberg,
Phys.\ Rev.\ Lett.\  {\bf 87}, 161602 (2001) [arXiv:hep-ph/0106295].

\bibitem{Giddings:2001bu}
S.~B.~Giddings and S.~Thomas,
Phys.\ Rev.\ D {\bf 65}, 056010 (2002) [arXiv:hep-ph/0106219].

\bibitem{Kaloper:2000jb}
  N.~Kaloper, J.~March-Russell, G.~D.~Starkman and M.~Trodden,
  Phys.\ Rev.\ Lett.\  {\bf 85}, 928 (2000)
  [arXiv:hep-ph/0002001].

\bibitem{Harris:2004xt}
C.~M.~Harris, M.~J.~Palmer, M.~A.~Parker, P.~Richardson,
A.~Sabetfakhri and B.~R.~Webber,
arXiv:hep-ph/0411022.
J.~Tanaka, T.~Yamamura, S.~Asai and J.~Kanzaki,
arXiv:hep-ph/0411095.

\bibitem{Kanti:2004nr}
P.~Kanti,
Int.\ J.\ Mod.\ Phys.\ A {\bf 19}, 4899 (2004)
[arXiv:hep-ph/0402168].

\bibitem{Giddings:2004xy}
S.~B.~Giddings and V.~S.~Rychkov,
Phys.\ Rev.\ D {\bf 70}, 104026 (2004) [arXiv:hep-th/0409131].
  H.~Yoshino and V.~S.~Rychkov,
  arXiv:hep-th/0503171.
  V.~Cardoso, E.~Berti and M.~Cavaglia,
  Class.\ Quant.\ Grav.\  {\bf 22}, L61 (2005)
  [arXiv:hep-ph/0505125].

\bibitem{Gregory:1993vy}
  R.~Gregory and R.~Laflamme,
  Phys.\ Rev.\ Lett.\  {\bf 70}, 2837 (1993)
  [arXiv:hep-th/9301052].

\bibitem{lovelock:498}
D.~Lovelock, J. Math. Phys.
  \textbf{12}, {498} {1971}.

\bibitem{Whitt:1988ax}
B.~Whitt,
Phys.\ Rev.\ D {\bf 38}, 3000 (1988).

\bibitem{Zwiebach:1985uq}
B.~Zwiebach,
Phys.\ Lett.\ B {\bf 156}, 315 (1985).
D.~G.~Boulware and S.~Deser,
Phys.\ Rev.\ Lett.\  {\bf 55}, 2656 (1985).
B.~Zumino,
Phys.\ Rept.\  {\bf 137}, 109 (1986).

\bibitem{hlr-progress} J.~L.~Hewett, B.~Lillie and T.~G.~Rizzo, in
preparation, 2005.

\bibitem{Rizzo:2005fz}
  T.~G.~Rizzo,
  arXiv:hep-ph/0503163.

\bibitem{Harris:2003db}
C.~M.~Harris, P.~Richardson and B.~R.~Webber,
JHEP {\bf 0308}, 033 (2003) [arXiv:hep-ph/0307305].

\bibitem{Sjostrand:2000wi}
  T.~Sjostrand, P.~Eden, C.~Friberg, L.~Lonnblad, G.~Miu, S.~Mrenna and E.~Norrbin,
  Comput.\ Phys.\ Commun.\  {\bf 135}, 238 (2001)
  [arXiv:hep-ph/0010017].

\bibitem{Anchordoqui:2003jr}
  L.~A.~Anchordoqui, J.~L.~Feng, H.~Goldberg and A.~D.~Shapere,
  Phys.\ Rev.\ D {\bf 68}, 104025 (2003)
  [arXiv:hep-ph/0307228].

\end{thebibliography}
\end{document}